# Seven steps to reliable cyclic voltammetry measurements for the determination of double layer capacitance

Dulce M. Morales[1] and Marcel Risch[1]

[1] Nachwuchsgruppe Gestaltung des Sauerstoffentwicklungsmechanismus, Helmholtz-Zentrum Berlin für Materialien und Energie GmbH, Hahn-Meitner-Platz 1, 14109 Berlin, Germany

E-mail: marcel.risch@helmholtz-berlin.de

**Abstract**

Discovery of electrocatalytic materials for high-performance energy conversion and storage applications relies on the adequate characterization of their intrinsic activity, which is currently hindered by the dearth of a protocol for consistent and precise determination of double layer capacitance ($C_{DL}$). Herein, we propose a seven-step method that aims to determine $C_{DL}$ reliably by scan rate-dependent cyclic voltammetry that considers aspects that strongly influence the outcome of the analysis, including (1) selection of a suitable measuring window, (2) the uncompensated resistance, (3) optimization of measuring settings, (4) data acquisition, (5) selection of data suitable for analysis, (6) extraction of the desired information, and (7) validation of the results. To illustrate the proposed method, two systems were studied: a resistor-capacitor electric circuit and a glassy carbon disk in an electrochemical cell. With these studies it is demonstrated that when any of the mentioned steps of the procedure are neglected, substantial deviations of the results are observed with misestimations as large as 61% in the case of the investigated electrochemical system. Moreover, we propose allometric regression as a more suitable model than linear regression for the determination of $C_{DL}$ for both the ideal and the non-ideal systems investigated. We stress the importance of assessing the accuracy of not only highly specialized electrochemical methods, but also of those that are well-known and commonly used as it is the case of the voltammetric methods. The procedure proposed herein is not limited to the determination of $C_{DL}$, but can be effectively applied to any other analysis that aims to deliver quantitative results via voltammetric methods, which is crucial for the study of kinetic and diffusion phenomena in electrochemical systems.

Keywords: double layer capacitance, cyclic voltammetry, benchmarking, electrocatalysis, fitting model

## 1. Introduction

A sustainable supply of clean, affordable energy relies not only on the efficient capture of energy originated from natural, inexhaustible sources such as sunlight or wind, but also on the possibility of its conversion and storage, so that it can be supplied and utilized when and where it is needed [1]. Diverse electrochemical systems, including water electrolyzers, metal-air batteries and fuel cells, are environmentally-friendly technologies that represent a potentially viable path towards sustainable energy supply. However, despite their large development during the last few years, these devices are not yet able to compete against fossil fuel-based technologies on an industrial scale due to their comparatively high energy input requirements, low efficiencies and high costs [2]. Hence, finding suitable electrode materials represents currently one of the main challenges in the field of electrocatalysis. This endeavor has led to the discovery of an extensive variety of materials over the past few decades. For that reason, refinement and standardization of the electrochemical methods used for their investigation is of utmost importance, not only for ensuring a proper assessment of their intrinsic catalytic properties, but also to enable a fair comparability between results obtained with different setups and/or in different laboratories. Guidelines and benchmark protocols for the evaluation of activity, selectivity and/or durability of electrocatalytic materials have been proposed in the framework of different electrochemical reactions, including those related to water electrolysis, and fuel cell technologies [3–6], as well as electrochemical processes of increasing environmental relevance such as the electroreduction of nitrogen [7–9] and of carbon dioxide [10–12]. Yet, a careful revision of how to adequately conduct and report electrochemical methods–as simple and well-know as cyclic or linear sweep voltammetries–is still necessary to achieve high-quality investigations. To illustrate the severity of this issue, we show herein the specific case of scan rate-dependent cyclic voltammetry, being a method recommended, though only briefly described, by various benchmark protocols to determine double layer capacitance of electrode surfaces.

Specific catalytic activity is typically reported in terms of current density, that is, current normalized by area unit. It has been broadly demonstrated that the geometric area of an electrode is an inadequate metric for normalization, as it introduces large errors by neglecting morphology, particle size and porosity of the investigated materials [13, 14]. An alternative parameter used for normalization of current is the surface area assessed by microscopic (e.g., atomic force microscopy) and gas adsorption techniques (e.g., nitrogen or hydrogen adsorption), in the cases of non-porous catalysts and of materials with high surface areas, respectively. However, such techniques are unable to discriminate electrically conductive components from non-conductive ones, as well as portions of the surface that are accessible to the electrolyte from those which are not [14–16]. Thereby, a more suitable parameter is the electrochemically active surface area (ECSA [=] $cm^2$), which allows to compare catalysts of different structural properties from an electrochemical standpoint. ECSA can be determined experimentally from the coulombic charge measured during characteristic adsorption/desorption reactions (e.g., hydrogen underpotential deposition or CO stripping), nonetheless, a major disadvantage of these methods is that they are only applicable to a limited number of materials, mainly metals [14, 17].

The ECSA can be determined from the double layer capacitance ($C_{DL}$ [=] µF) and the specific capacitance of any investigated electrode material ($C_S$ [=] µF $cm^{-2}$) according to **Equation 1** [16].

$$\text{ECSA} = \frac{C_{DL}}{C_S} \qquad (\text{Eq. 1})$$

Determination of $C_S$ has been identified as one of the main challenges for the accurate assessment of ECSA [14]. Since its value is typically unknown, it has become a common practice to use a single $C_S$ value to obtain ECSA for any material independently of its nature, evidently leading to largely under or overestimated ECSA values [15]. Yet, with $C_S$ established as the main source of inaccuracy for ECSA, the importance of the accurate determination of $C_{DL}$ is generally overlooked. Moreover, even if $C_S$ is not known relative changes in ECSA can be observed by monitoring accurately $C_{DL}$, and it has been recently proposed that, instead of using ECSA, the catalytic activity could be reported in terms of net measured current along with the current normalized by geometric area and by $C_{DL}$ [18]. Such an approach, though managing to circumvent the problematic need for $C_S$, still requires knowing $C_{DL}$ accurately to ensure its reliability and, thus, comparability between different catalytic materials. In this context, we propose herein a seven-step methodology for the determination of $C_{DL}$ by scan rate-dependent cyclic voltammetry that allows, firstly, to identify optimal measurement parameters for high-quality data collection, and secondly, to perform a high-quality data analysis to ensure high reliability. We demonstrate the proposed methodology and possible pitfalls at each step by means of two different systems: an electrical circuit where the investigated element is a capacitor, and an electrochemical cell where the investigated surface is a smooth glassy carbon disk immersed in 0.1 M NaOH solution.

## 2. Methods

Two different setups were used for the electric/electrochemical experiments. The first setup consisted of a resistor-capacitor electric circuit (RC-circuit) built with two 100 Ω metal film resistors (± 1% tolerance, Yageo) connected in parallel for a total resistance of 50 Ω. The resistors were soldered in series with two 10 µF solid tantalum resin dipped radial capacitors (± 10% tolerance, rated voltage 35 V, AVX) parallel to each other, for a total capacitance of 20 µF. The circuit was kept in a Faraday cage during the measurements. The second setup was a three-electrode configuration single-compartment electrochemical cell (EC-cell) made from polymethyl pentene, where a glassy carbon electrode of 4 mm diameter was used as the working electrode. Prior to the measurements, the working electrode was polished with 0.05 µm alumina paste (ALS Inc.) and rinsed with water and ethanol. A Hg|HgO electrode (ALS Inc.) and a coiled platinum wire were used as the reference and the counter electrodes, respectively. The measurements were conducted in argon-purged 0.1 M NaOH aqueous solution (Sigma-Aldrich) as electrolyte. The EC-cell was kept inside a KB240 climate chamber (Binder) to ensure a controlled temperature of 25 °C during the measurements. For both setups, namely the RC-circuit and the EC-cell, a Reference 600+ potentiostat (Gamry) was used to control the applied potentials. The offsets of the current ranges and cable capacitance were corrected according to the manufacturer's calibration procedure. Cyclic voltammograms (CVs) were recorded within selected potential windows using scan rate values in the range from 0.01 to 10 V s$^{-1}$. Unless indicated differently, the data was collected by sampling during the entire voltage step of the digital staircase that approximates an analog ramp ("surface sampling mode") using a step size of 0.2 mV, optimized current limit, and optimized dynamic iR-drop compensation (positive feedback). In the case of the EC-cell, 5 consecutive CVs were recorded for each experiment, and the measured current values corresponding to double layer charging were extracted solely from the last cycle. Data obtained by cyclic voltammetry was processed with the software Origin 9.5 (OriginLab Corporation), using either linear or allometric regression models to fit data from current vs scan rate plots. Electrochemical impedance spectroscopy (EIS) was conducted in the frequency range from 100 to 0.05 kHz with 10 mV (RMS) amplitude at open circuit. EIS spectra were analyzed and/or simulated using the software Echem Analyst (Gamry).

## 3. Determination of double layer capacitance

When an electrode surface is subjected to a voltage ramp, a steady-state capacitive current is observed in a short time ($i_C$) if the only process taking place within the voltage range is the charging of the double layer, that is, movement of ions on either side of an electrode/electrolyte interface. For ideal capacitors, $i_C$ is related to the capacitance (C) and to the scan rate (ν) as described in **Equation 2** [19].

$$i_C = \nu \cdot C \qquad (Eq.\ 2)$$

A common method for obtaining the double layer capacitance ($C_{DL}$) consists in recording cyclic voltammograms at various scan rates within a potential region where no redox processes take place, extracting $i_C$ from the anodic and cathodic scans of the recorded voltammograms, and subsequently extracting $C_{DL}$ from the slope of the resulting $i_C$ vs ν plot. $C_{DL}$ can also be obtained by means of electrochemical impedance spectroscopy (EIS), provided that a suitable equivalent circuit was selected [19, 20]. However, we focus herein in its determination by cyclic voltammetry, stressing its advantage as a more widely available method than EIS.

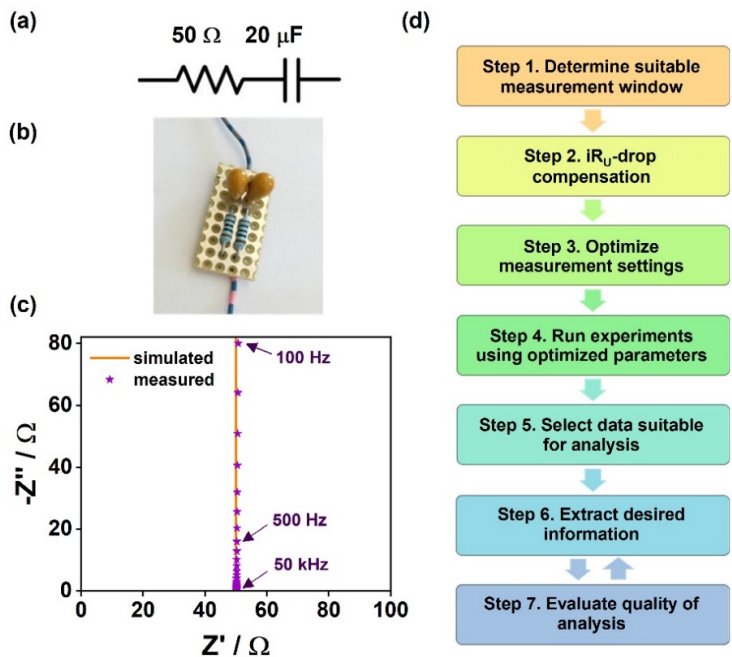

**Figure 1**. (a) Electrical diagram and (b) photograph corresponding to the RC-circuit used in this work. (c) Nyquist plot of the built RC-circuit obtained by electrochemical impedance spectroscopy at open circuit voltage with an amplitude of 10 mV (RMS), compared with a simulated ideal equivalent circuit

(50 Ω resistance in series with 20 μF capacitor). (d) Seven-step procedure for ensuring reliable cyclic voltammetry-based measurements.

Since our purpose was to establish a reliable methodology for the careful determination of double layer capacitance *via* cyclic voltammetry, it was necessary to study a system that delivers a predictable outcome. The selected system was an electric circuit consisting of standard electric elements with a total capacitance of 20 μF in series with 50 Ω resistance (**Figure 1a** and **b**). Such a system, hereafter denoted as RC-circuit, was convenient because on the one hand, the capacitance value is known, and on the other hand, the circuit is expected to exhibit a behavior close to ideal, thus allowing us to systematically observe the different steps of the experimental procedure and data analysis in relation to the outcome of the analysis. A comparison between the Nyquist plot of the built RC-circuit obtained by electrochemical impedance spectroscopy, and a simulated plot for the equivalent ideal circuit is shown in **Figure 1c**, demonstrating that the response of the RC-circuit is indeed close to ideal.

Being aware that electrochemical systems generally exhibit considerably more complex behaviors than ideal electric circuits [21, 22], we selected and investigated a second system which involved an electrode/electrolyte interface. This system, hereafter denoted as EC-cell, consisted of an electrochemical cell where a polished glassy carbon disk immersed in 0.1 M NaOH aqueous solution was the subject of study. With the aid of the two selected systems, namely the RC-circuit and the EC-cell, we were able to identify, firstly, sources of error that are easy to overlook, and secondly, strategies to circumvent them, integrating these strategies in the seven-step protocol illustrated in **Figure 1d**. In the following subsections, each of the seven steps is described and discussed, and their importance in terms of their impact on the outcome of the analysis is clearly illustrated with diverse measurements conducted using the two selected model systems.

*Step 1: Determine suitable measurement window*

When conducting cyclic voltammetry, two main parameters are to be chosen: (i) the measurement window, which is set by the lower and upper potential limits of the voltammogram, and (ii) the speed at which the potential is varied within these two limits, that is, the scan rate. The first step of the procedure focuses on the first of the two parameters. The second parameter will be discussed in detail later on (Step 5).

For the determination of $C_{DL}$, it is necessary to ensure that the data is recorded within a potential region where the charging of the double layer is the main contribution to the measured net current while current contributions originated from other electrochemical processes, including redox reactions and adsorption/desorption processes, are negligible. This potential region can be identified by recording a voltammogram in a relatively wide potential range. It is recommended to use a slow scan rate between 1 and 10 mV s$^{-1}$ in order to minimize capacitive contributions (Equation 2), thus facilitating the observation of redox processes. Note, however, that redox processes become less intense as the scan rate decreases, which on the one hand makes it difficult to distinguish them and on the other hand emphasizes measurement artifacts more prominently [23]. In addition to this, it is advised to conduct this experiment after recording consecutive CVs until a steady response is observed as a preconditioning step. In the case of the RC-circuit, virtually any potential window can be used for the determination of its capacitance as it exhibits the rectangular profile expected for an ideal capacitor [24] throughout the entire window, limited only by the potentiostat capabilities (**Figure 2a**). In contrast, for electrochemical systems, the potential range suitable for the measurement depends on the nature of the electrode and electrolyte used, and corresponds to the window where neither the electrode nor the electrolyte undergoes electrochemical processes different to double layer charging. For aqueous systems, the cathodic and anodic limits of the potential region where the electrolyte does not undergo electrochemical reactions are set by the hydrogen evolution (HER) and the oxygen evolution (OER) reactions, respectively, displaying a stability window of at least 1.23 V. This window can be enlarged by use of "water-in-salt"-type electrolytes to achieve stability windows as large as 3 V [25, 26]. However, the potential window suitable for determining double layer capacitance is shortened considerably by the presence of any species in the electrolyte that may undergo redox processes at the electrode/electrolyte interface [27], e.g. reduction of dissolved oxygen (ORR), or adsorption/desorption of ions in the solution as shown in **Figure 2b**. Furthermore, species present at the electrode surface that undergo redox processes within the electrolyte stability window will shorten further the potential range that is suitable for the determination of $C_{DL}$. In the case of the electrochemical system investigated here, recording a wide CV allowed us to identify a potential window of about 600 mV as the maximum measurement window for $C_{DL}$ determination in both the presence and the absence of air, as shown in Figure 2b.

A widespread practice for selecting the upper and lower potential limits for the measurements consists of using a narrow potential window [16], typically of 100 mV [4], centered at the open circuit potential (OCP). However, following this approach uncritically may lead to large deviations. Firstly, while it is generally assumed that the OCP falls within the potential region where no redox processes take place due to the fact that no external current is flowing through the system at this potential, there are cases where long times are required to reach stable OCP values after cell assembly [28]. Recording a wide CV similar to the one shown in Figure 1b results helpful for assessing the suitability of the OCP as the center of the measurement window. A second deviation originates from the standardized use of the 100 mV measurement window because it is necessary that a steady charging current is reached within the investigated voltage window. This is described by the transient period, namely, the period in which double layer charging has not yet reached steady state when transitioning from an anodic to a cathodic sweep (or *vice versa*), whose duration depends on the properties of the system. If the transient period is short, a window of–or even narrower than–100 mV might be wide enough for allowing the system to reach a steady charging current. That is the case of the RC-circuit as shown in **Figure 2c**, where both steady currents, anodic and cathodic, clearly overlap within the different measurement windows shown. However, this is often not the case for electrochemical systems, as other factors related to intrinsic irreversibility of charging/discharging processes [24], and the presence and transport of ions at the interface [27, 29, 30], as well as electrode conductivity [22, 27, 29] and cell resistance (discussed more in detail in Step 2) could impact substantially the initial stages of the double layer charging. Moreover, if the transient periods observed during the anodic CV scans are different to those of the cathodic CV scans and the measuring window is too narrow, the $C_{DL}$ values obtained from the corresponding data set could also deviate substantially from each other. Thus, it is often necessary to select measuring windows larger than 100 mV not only to ensure reaching a steady charging current, but also to ensure that $i_C$ values extracted from the anodic and cathodic scans are similar. **Figure 2d** illustrates this issue for the case of the EC-cell, showing that a 100 mV window is too narrow for this system as clear overlaps of both the anodic and cathodic currents are observed only for CVs recorded with 500 and 550 mV measuring windows. Furthermore, comparing the currents recorded within different measuring windows, larger differences were observed for the cathodic than for the anodic scans, suggesting that the transient periods of the former are considerably larger than for the latter. It is thus advisable to select the suitable measurement window by recording CVs at the same scan rate within potential ranges of increasing size, until a clear overlap of both the anodic and cathodic steady current is observed. For the systems studied here, namely the RC-circuit and the EC-cell, measurement windows of 100 and 500 mV, respectively, centered at 0 V and at OCP, respectively, were selected for the subsequent experiments, unless indicated differently.

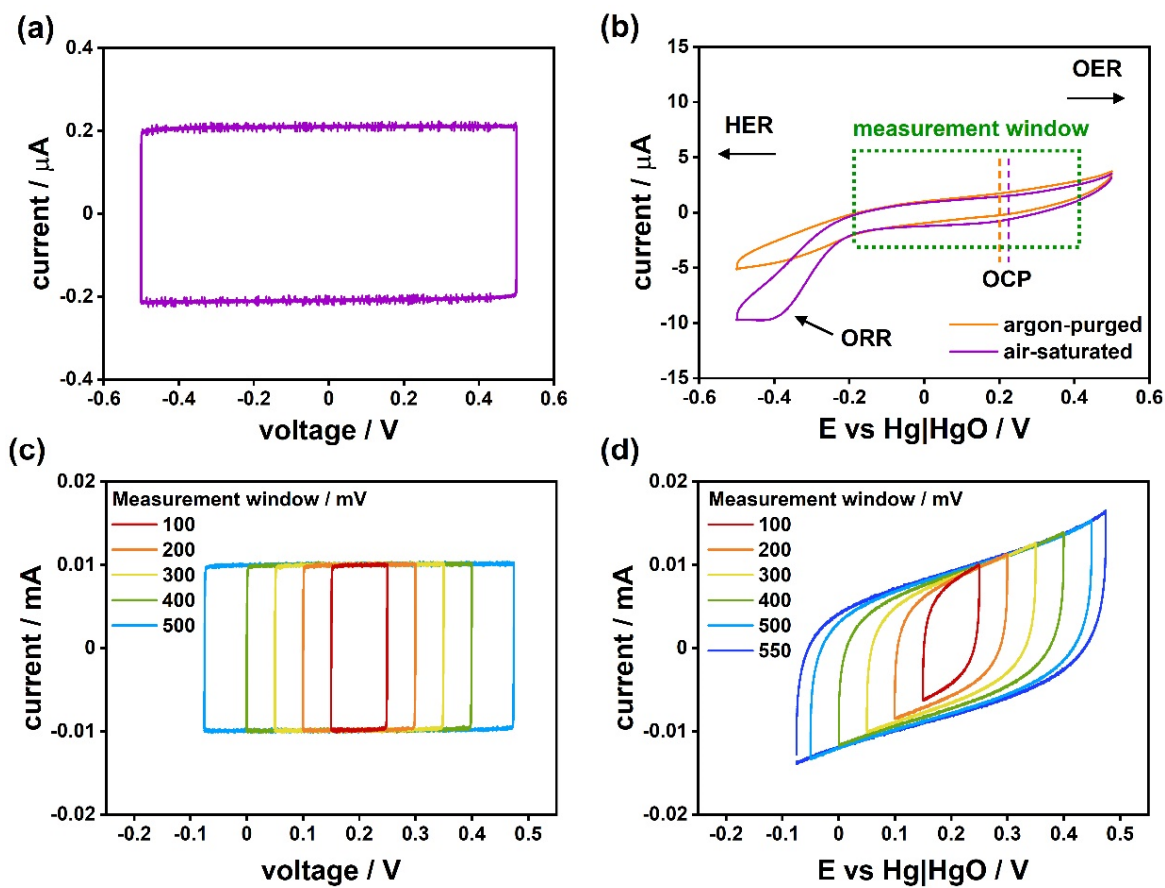

**Figure 2**. Cyclic voltammograms recorded for (a,c) the RC-circuit and (b,d) the EC-cell, using scan rates of (a,b) 10 mV s$^{-1}$ and (c,d) 500 mV s$^{-1}$ within different measurement windows. Note that (a,b) and (c,d) are displayed with the same X-axis scale to facilitate the comparison.

Note that the electrode surface of the EC-cell studied here was made of glassy carbon, whose inert nature allows for conducting these measurements in a large potential range without the interference of redox processes. However, this may not be the case for actual electrocatalytic materials. For example, Co, Ni or Fe-based compounds, which have been studied as catalysts for diverse reactions in aqueous media, display redox peaks within the measurement window shown in Figure 2b [31, 32]. The presence of redox peaks may not only make difficult to find a sufficiently large CV window for the determination of $C_{DL}$, but in some cases it may even be impossible, specially in those involving more than one interfering redox peak [33]. It is important to keep in mind that the determination of $C_{DL}$ via scan-dependent cyclic voltammetry should not be regarded as a universal technique, and a careful evaluation of its suitability when applied to a given electrochemical system is necessary.

**Summary of Step 1**: Record a CV at a slow scan rate to identify a redox process-free potential window and select a window within that ensures that a steady charging current is reached and that $i_C$ values extracted from the anodic and cathodic scans are similar.

*Step 2: $iR_U$-drop compensation*

In three-electrode configuration cells, the electrode of interest (the working electrode) is investigated with respect to the reference electrode. A potentiostat controls the voltage between them by injecting current through the counter electrode. The response observed at the working electrode depends not only on the nature of the electrode/electrolyte interface and on the experiment to be conducted, but also on parameters related to the electrochemical setup, including cell geometry, electrode shape, solution conductance, wiring, among others. The potentiostat is able to compensate for the resistance encountered between the working and the counter electrodes in order to achieve the desired voltage or current at the working electrode. However, it is not able to compensate the resistances encountered between the working electrode and the equipotential surface at the tip of the reference electrode, leading to an unavoidable difference between the desired potential and the actual potential applied at the surface of the working electrode [34]. The sum of such resistances is called uncompensated resistance ($R_U$) and the difference in potential is commonly known as $iR_U$-drop, as it depends on both $R_U$ and the current (i) flowing through the system, according to Ohm's law [19]. Thereby, the $iR_U$-drop must be considered for every attempt of quantitative analysis involving data collected with 3-electrode configuration cells.

Compensation of the $iR_U$-drop is of special importance for the specific case of the determination of $C_{DL}$ by voltammetric methods, as the uncompensated $iR_U$-drop leads to non-linear potential sweeps, and consequently, to a misread of the voltammetric data [35]. For that reason, the second step of the procedure focuses on the compensation of this potential drop. Different experimental methods, such as EIS, positive feedback or current interrupt, allow for the determination of $R_U$ and their advantages (and disadvantages) have been extensively discussed [6, 36–38]. Once $R_U$ is determined, the measured potentials can be corrected– or compensated– according to **Equation 3**. Data for the determination of $C_{DL}$ should be extracted from the corrected voltammograms.

$$E_{corrected} = E_{measured} - iR_U \qquad (Eq. 3)$$

Most modern potentiostats offer the possibility of conducting iR-drop compensation dynamically, that is, while the measurement is running. This option ensures that the desired potentials are applied independently of the current flow and further offers the advantage of simplifying the post analysis, as the step for correcting the measured potentials is no longer needed. It requires that $R_U$ is known prior to the measurement, since its value has to be input for the potentiostat to conduct dynamic compensation. However, $R_U$ cannot be fully compensated in this manner, as it can lead to potentiostat oscillation. While it is generally recommended to input 80-90% of $R_U$ for dynamic compensation, it is advisable to find an optimum input value, namely, as large as possible without causing oscillation. **Figure 3** shows the upper edge of voltammograms of the RC-circuit and the EC-cell recorded within the same measurement windows using different resistance values as input for dynamic compensation. Note that the RC-circuit was investigated analogously to a two-electrode cell, where $R_U$=0 and thus, $iR_U$-drop compensation is not needed. In this case, the input resistance values were calculated with respect to the total resistance of the circuit (50 Ω) to illustrate the effect of overcompensation.

The impact of compensation is stronger in the potential region where the current is rising than it is in the region where a steady current is reached, indicating that compensation of $iR_U$-drop is of even higher relevance for interfaces with long transient periods, specially if narrow measurement windows were selected. On the other hand, oscillation due to overcompensation can clearly introduce large deviations and must be avoided. In fact, overcompensation led to a potential shift in a direction opposite to the shift observed from (under)compensated data, as indicated by the arrows in Figure 3. The upper edge of the CVs recorded for the EC-cell shifted to lower potentials as the resistance value input for compensation increased, however, as the data underwent overcompensation, the edge shifted back to larger potentials. This observation demonstrates the importance of optimizing the dynamic $iR_U$-drop compensation prior to conducting quantitative analyses.

**Summary of Step 2**: Determine the uncompensated resistance and use it to correct the measured potentials. If compensation is done dynamically, make sure that the input value of resistance is as large as possible without causing oscillation.

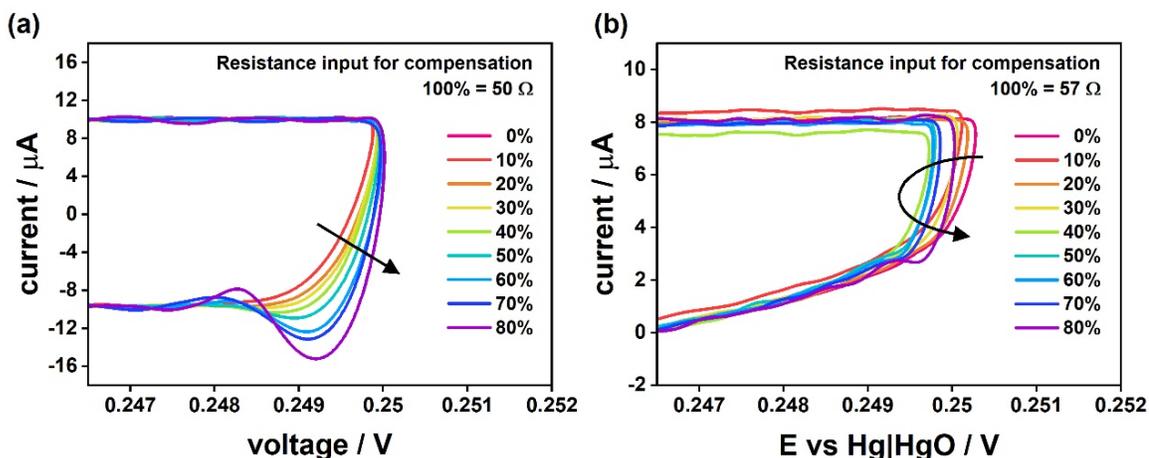

**Figure 3.** Upper edge of cyclic voltammograms of (a) RC-circuit and (b) EC-cell recorded at 0.5 V s$^{-1}$ scan rate, using different resistance values as input for dynamic iR$_U$-drop compensation. Arrows indicate the trend of the recorded voltammograms with respect to an increasing resistance value input for compensation.

*Step 3: Optimize measuring settings*

Most modern potentiostats integrate a series of optional settings that allow for the enhancement of their capabilities towards specific purposes. Though this can be quite convenient, it is important to keep in mind its main implication: different settings may lead to different measurement outcomes. It is therefore necessary to fully understand the potentiostat functionalities in order to select adequately the settings for the experiment. Manuals as well as diverse application notes have been made available by different potentiostat manufacturers for this purpose. Yet, it is advisable to make use of dummy cells for testing and understanding the different settings, as they allow a direct observation of the potentiostat responses to them, as well as an assessment of whether maintenance or calibration of the potentiostat is needed. Thus, the third step of the procedure involves the careful selection of potentiostat-dependent settings to ensure that the experimental data is collected adequately.

To illustrate the importance of optimizing the measuring settings, CVs of the RC-circuit were recorded varying three selected settings available for cyclic voltammetry in Gamry potentiostats: the current range, the size of the voltage step of the digital staircase approximating an analog ramp, and the sampling mode, namely, the mode in which data is sampled throughout the voltage step of the digital staircase (**Figure S1**). Note that these settings are often available in potentiostats from other manufacturers, and that there may be other settings not considered here that could also have a substantial impact on the voltammetric responses. **Figure 4a** shows CVs obtained at a scan rate of 0.5 V s$^{-1}$ with maximum currents set at either fixed current values of 10, 100, and 1000 µA or in the automatic current range mode. The current expected for an ideal capacitor of 20 µF at this scan rate is 10 µA according to Equation 2. Noisier responses were observed for the cases where the difference between this value and the selected maximum current was larger. Moreover, in addition to higher noise, setting larger maximum currents led to offsets with respect to the expected current towards lower values (more positive in the case of cathodic scans) as shown in the inset of Figure 4a. While a maximum current set to 100 µA led only to a small deviation, 1000 µA (hundred times larger than the expected current) led to an offset in current of about 20%. Interestingly, the option for adjusting the current range automatically, though convenient, is not as precise as selecting 10 µA as maximum current input. Furthermore, this option may introduce current spikes as the current range is (automatically) adjusted. Thus, it is recommended to input an optimal fixed current range rather than using automatic adjustment for quantitative analyses. In the case of the determination of C$_{DL}$ it is advisable to optimize the current range for each scan rate tested.

**Figure 4b** shows CVs obtained at a scan rate of 0.5 V s$^{-1}$ with potential steps of the digital staircase of different sizes. The potential step size defines how many potential points are to be collected, delivering higher resolution with smaller step sizes. Where data storage is not of concern, it is recommended to collect larger amount of data points for more accurate analysis. For the case of the determination of C$_{DL}$ by cyclic voltammetry, the step size becomes relevant in two cases. The first case is when the measurement window is narrow, since using large potential step sizes narrows further the data collected. The second case is related to the third setting parameter, namely the sampling mode, which can be conducted in two limits: (i) by sampling the entire staircase step in its full duration, or (ii) by only sampling data at the end of it (**Figure S1**). The latter mode is useful for experiments where it is desired to minimize capacitive contributions as it allows for the capacitive current to decay before acquiring the data. Thereby, if used for investigation of $C_{DL}$, it will lead to large underestimations of the capacitive current, specially in the case where a large potential step size was used. A comparison of these two modes is shown in **Figure 4c** and **4d**, illustrating clearly that the two modes deliver substantially different responses (note that for these two figures the scale is identical).

**Summary of Step 3**: Understand fully your potentiostat capabilities and identify optimal measuring settings.

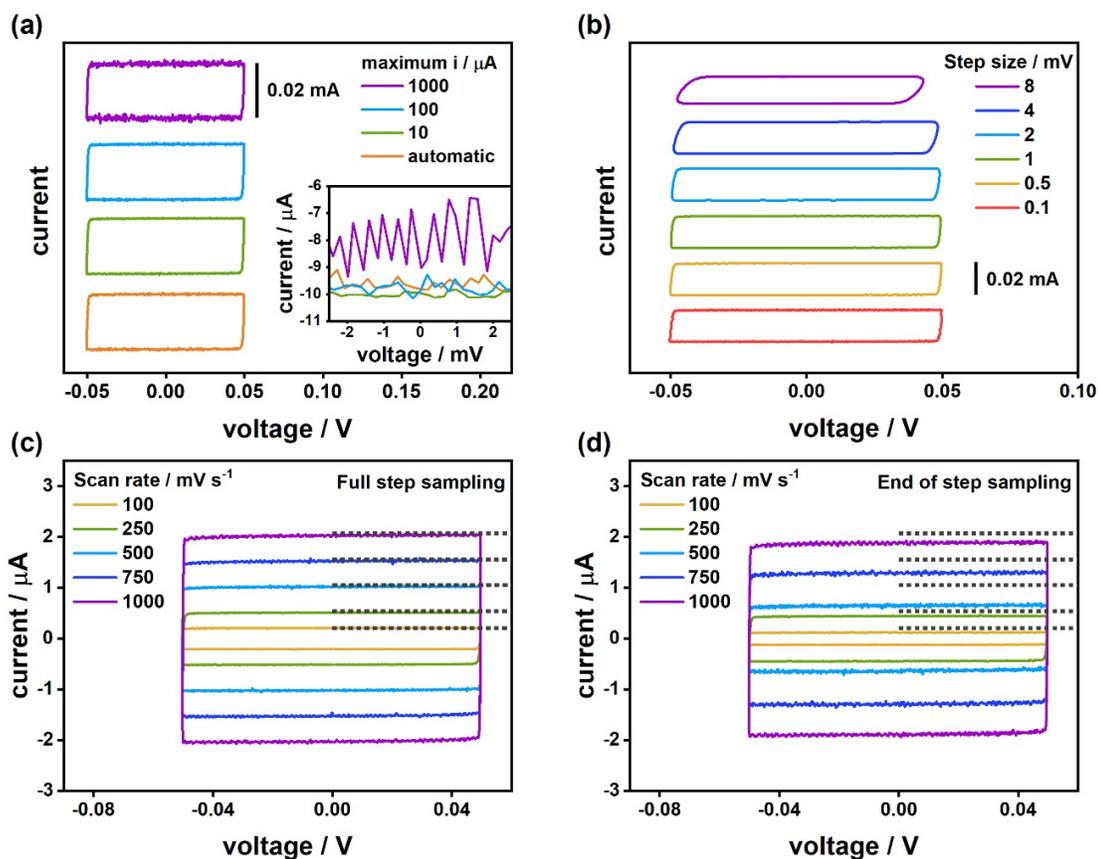

**Figure 4**. Cyclic voltammograms of RC-circuit comparing settings available in Gamry potentiostats for data collection including (a) maximum current (0.5 V s$^{-1}$ scan rate, step size 0.2 mV, sampled over the full step), (b) step size (0.5 V s$^{-1}$ scan rate, current limit 0.01 mA, sampled over the full step), and sampling modes (c) over the full step and (d) at the end of the step, at different scan rates (optimized current limit, step size 0.2 mV). The inset of (a) shows the currents measured in a 4 mV window during the cathodic scan with different settings of maximum current. Y-axis scales of plots (c) and (d) are identical with dashed lines set at the same Y-axis position to facilitate comparisons between the two plots.

*Step 4: Run experiments using optimized parameters*

The first three steps of the procedure focus on selecting the parameters that are suitable for conducting a given measurement with a given electrochemical system, meaning that these parameters could be different for different electrode/electrolyte interfaces. It is only in Step 4 when the actually necessary data for the analysis is acquired. The experiment consists of recording CVs within the selected measurement window (Step 1) at different scan rates, using the optimized settings (Steps 2 and 3) for data acquisition. It is recommended to use a scan rate range as wide as possible, avoiding on the one hand too slow scan rates that make the experiment unnecessarily time-consuming, and on the other hand too large scan rates that may lead to large deviations originating from potentiostat limitations (e.g., insufficient sampling frequency). These limitations can be identified by means of a dummy cell or an electric circuit similar to the RC-circuit, since for either case the current measured during a CV can be predicted, thus facilitating the observation of deviations at high scan rates. In addition to this, it is advisable to record not one but several consecutive CVs until unchanging responses are observed. The $i_C$ data is later extracted solely from the last CV.

The difference between using non-optimized and optimized settings for data acquisition is illustrated by **Figure 5a** and **5b**, respectively, showing the corresponding voltammetric responses of the RC-circuit at different scan rates. Let us take as an example a scan rate of 10 V s$^{-1}$. The current expected, namely 0.2 mA according to Equation 2, is indicated by a dotted line. Clearly, this current was reached only when the measurement was conducted using optimized settings (Figure 5b).

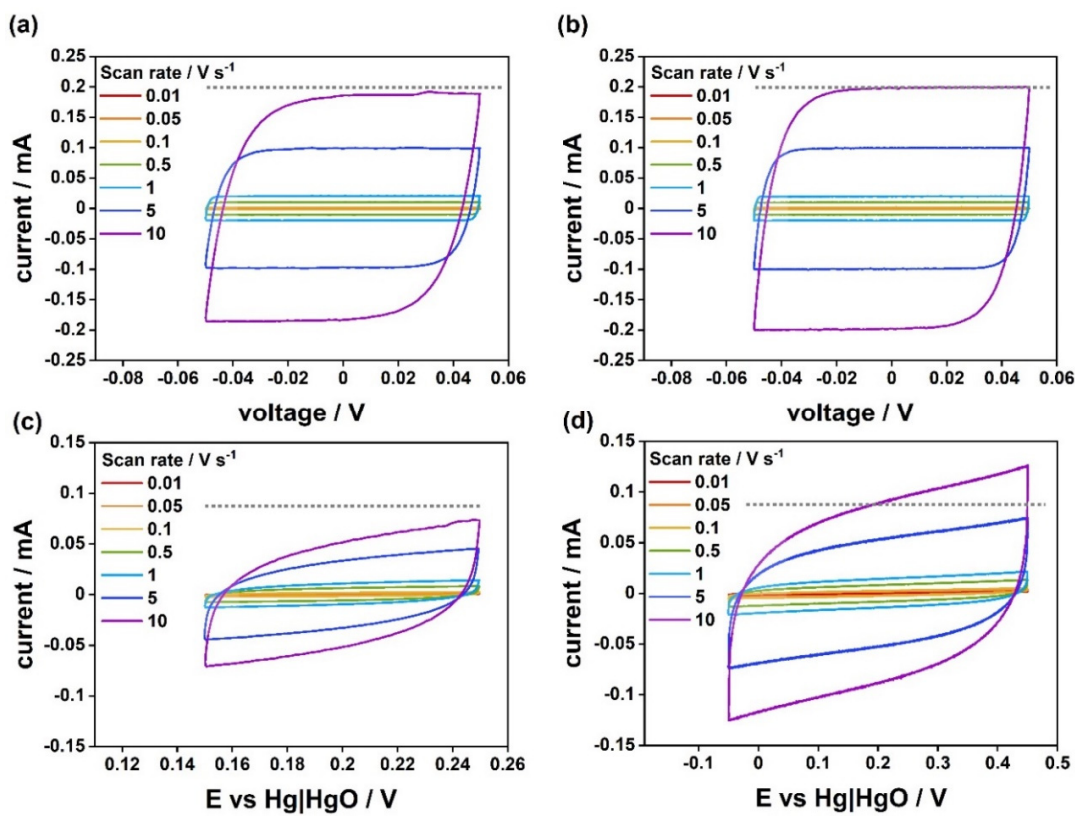

**Figure 5**. Cyclic voltammograms of (a,b) RC-circuit and (c,d) EC-cell at different scan rates using (a,c) non-optimized settings (100 mV measurement window, no $iR_U$-drop compensation, automatic current limit, step size 1 mV) and at (b,d) optimized settings (50% $iR_U$-drop compensation for the EC-cell, current limit optimized for each scan rate, step size 0.1 mV). Figures at the left are displayed with the same scales as those at the right. Dot lines in Figures at the left are set in the same Y-axis position as in those at the right to facilitate comparisons between plots.

While an underestimation of the capacitive current is easy to identify for the RC-circuit as its capacitance is known and its behavior predictable, it can be easily overlooked for electrochemical systems. Such is the case of the EC-cell, as shown in **Figures 5c** and **5d** obtained with non-optimized and optimized settings, respectively. The dotted line in these two figures is situated at a current value of 87 µA which corresponds to the current extracted at a potential of 0.2 V vs Hg|HgO from the CV recorded at 10 V s$^{-1}$ using optimized settings (**Figure**

**5d**). Meanwhile, the current extracted at the same potential from a CV recorded at the same scan rate, for which optimization of measurement settings was not done (**Figure 5c**) was about 52 μA, representing an underestimation of $i_C$ of about 40% with respect to the former case. It is evident that a substantial underestimation of capacitive current is the consequence of not using the optimal settings, which could lead to a substantial misestimation of the $C_{DL}$. The impact of the underestimations observed for the RC-circuit and EC-cell on the $C_{DL}$ value obtained after data analysis will be discussed in detail later on (Step 6).

**Summary of Step 4**: Record continuous CVs in an optimal potential window using optimized measuring settings and at different scan rates in a wide scan rate range.

*Step 5: Select data suitable for analysis*

The voltammetric data required for the analysis has been acquired during Step 4, however, that does not necessarily mean that all the collected data will be useful. Step 5 of the procedure focuses on identifying the data that is suitable for analysis. In the case of the determination of $C_{DL}$, the data corresponding to capacitive current is extracted from the anodic and cathodic sweeps of the–last if several–CVs recorded at different scan rates. In the case of the RC-circuit and the EC-cell, the charging current data was extracted at the potentials at which the corresponding CVs were centered, namely, 0 and 0.201 V (OCP), respectively. During the fifth step of the procedure, the scan rate range is narrowed by taking into consideration (i) possible non-capacitive processes contributing to the measured current, (ii) possible ion transport limitations, and (iii) the similarity between data extracted from the anodic and the cathodic scans, all of which depend on the specific characteristics of the electrochemical system investigated. For the first consideration, it should be noted that competing faradaic processes contribute more significantly to the overall measured current at lower scan rates, whereas larger contributions of the capacitive currents are achieved at higher scan rates. This is illustrated in **Figure 6a**, where simulated current vs scan rate plots corresponding to an ideal RC-circuit (50 Ω in series with 20 μF) and to an ideal Randles circuit (50 Ω in series with 20 μF with a parallel resistance of 1 MΩ) are shown. The data corresponding to the latter was simulated according to **Equation 4** (see discussion in supporting information). The parallel resistor ($R_P$) represents the effect of resistances encountered at the electrode/electrolyte interface. It contributes to the total resistance ($R_T$), and together with the voltage applied (V), to the net current ($i_{net}$) according to Ohm's Law. Note that Equation 4 cannot be used as a valid model for electrochemical systems where $R_P$ is potential-dependent. Indeed, although the capacitance is the same for both systems, the measured currents differ from each other more substantially the lower the scan rate is, while the contribution of $R_P$ becomes negligible at larger scan rates. Thereby, in cases where contributions of faradaic processes are difficult to avoid, for instance due to the presence of redox species in the electrochemical system with redox potentials near to the measurement window, data measured at slow scan rates should be discarded.

$$i_{net} = \frac{V}{R_T} + \nu \cdot C \qquad \text{(Eq. 4)}$$

The second consideration is related to ion transport limitations and is described in detail in a report by Wang and Pilon [29]. In summary, using too large scan rates may result in an underestimation of capacitive currents in cases of systems where ion transport is limited either by low diffusion capabilities of the ions in the electrolyte (due to low diffusion coefficient and/or to solvation effects), or by the use of electrodes with mesoporous structures, or by both. Additionally, large electrolyte concentrations as well as poor electrode conductivity may also lead to lower currents. All these limitations may result in a decrease of the slope in the $i_C$ vs ν plot, leading subsequently to an underestimation of $C_{DL}$ (Step 6). Thus, selecting an appropriate scan rate range, though often neglected, is not trivial and should be considered in accordance with the characteristics of the specific electrochemical system under investigation.

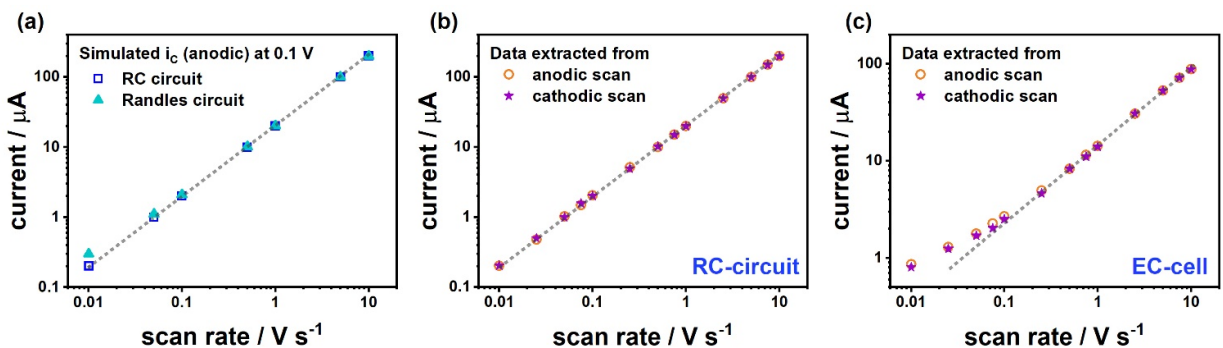

**Figure 6**. (a) Anodic current as function of scan rate simulated at a voltage of 0.1 V for an ideal RC circuit (50 Ω in series with 20 μF) and for an ideal Randles circuit (50 Ω in series with 20 μF with a parallel resistance of 1 MΩ). Absolute net currents extracted from anodic and cathodic voltammetric sweeps of (b) RC-circuit and (c) EC-cell at a voltage of 0 V and at an electrode potential of 0.2 V vs Hg|HgO, respectively, recorded at various scan rates. Note that the X and Y axis are shown in logarithmic scale. Dotted lines are included to guide the eye.

The third consideration involves a comparison of the data collected from the anodic and the cathodic voltammetric scans. As discussed previously (Step 1), for an ideal capacitor the anodic and cathodic capacitive currents are equal, only differing in the sign which, by convention, is positive for anodic current and negative for cathodic current. Thus, provided that the potentiostat is properly calibrated, the absolute $i_C$ values are expected to overlap with each other in a current vs scan rate plot, as shown in **Figure 6b** for the RC-circuit. Note that the overlap occurs for all data points shown in Figure 6b, each of which corresponds to a different scan rate value. Nevertheless, this is not the case for the EC-cell as shown in **Figure 6c**, where a fair overlap of the cathodic and anodic currents was observed at scan rates above 0.5 V s$^{-1}$, while at lower scan rates the value of the two (absolute) currents differ. Moreover, the resulting plot resembles that of the Randles circuit shown in **Figure 6a**, indicating that the system does not behave as an ideal capacitor (this will be discussed in detail in Step 7). Thus, it is convenient to select the data set corresponding to large scan rates to ensure that the majority of current contributions originates from capacitance. As the investigated electrode surface is that of a smooth glassy carbon disk immersed in NaOH solution of moderate concentration (0.1 M), ion transport is not expected to be limiting for the larger scan rates of the investigated range. For our examples in **Figure 6**, we selected the scan rates from 0.01 to 10 V s$^{-1}$ for the RC-circuit and from 0.5 to 10 V s$^{-1}$ for the EC-cell.

**Summary of Step 5**: Extract charging currents recorded during the last anodic and cathodic sweeps for each scan rate and plot their absolute values against the scan rate, both in logarithmic scales. Discard data points that deviate substantially from the linear current-scan rate relation.

*Step 6: Extract desired information*

After selecting a suitable data set, data processing takes place and the desired information is extracted, which is the focus of Step 6. In the case of the determination of $C_{DL}$, the capacitance is extracted from the current vs scan rate plot by fitting the data to a model. The most common model used is that of the ideal capacitor, in which case a linear regression is used to obtain $C_{DL}$ from the slope according to **Equation 5**. If the intercept is set to zero for the fitting, the equation obtained from the linear fit is equal to that of the ideal capacitor (Equation 2), with the variables Y and X being the current and the scan rate, respectively, and the slope b being the capacitance. The validity of this model will be discussed in detail later on (Step 7).

$$Y = bX \qquad (Eq.\ 5)$$

**Figure 7a** shows the two current vs scan rate data sets collected for the RC-circuit extracted from the CVs shown in **Figure 5a and 5b** (corresponding to non-optimized and optimized measurement settings, respectively) with their corresponding linear fits. The average of the capacitances determined from the anodic and the cathodic scans were 19.8 and 18.8 μF for the optimized and non-optimized data sets, respectively. Considering that the circuit integrated a capacitor of 20 μF with ± 10% tolerance, both values fell within the expected capacitance window, with deviations of 1% and 6% with respect to the nominal capacitance of 20 μF, respectively.

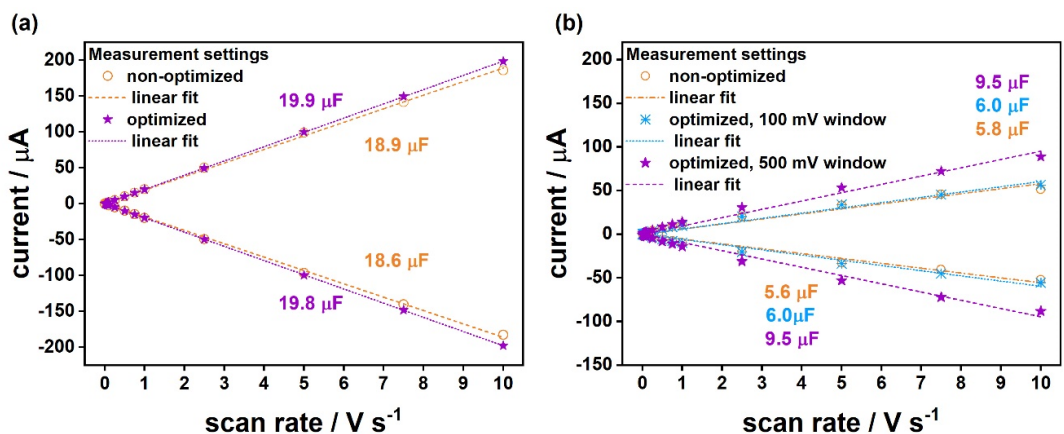

**Figure 7.** Current as a function of scan rate of (a) RC-circuit obtained with optimized and non-optimized measurement settings, and of (b) EC-cell, obtained with optimized and non-optimized measurement settings, and comparing measurement windows of 100 and 500 mV. Dot lines correspond to corresponding linear regression fits. The resulting capacitance values are shown with the colors corresponding to their respective data set.

In the previous section (Step 5), it was stablished that the EC-cell does not exhibit the behavior of an ideal capacitor. Yet, the linear model was used here to fit the data corresponding to this system to be able to observe the outcome of the analysis and to compare it later on with a more suitable model (Step 7). Data sets of current vs scan rate collected with the EC-cell are shown in **Figure 7b** together with their corresponding linear fits. The difference between using optimized and non-optimized measurement settings was more dramatic for the EC-cell than it was in the case of the RC-circuit, obtaining $C_{DL}$ values of 9.5 and 5.8 µF, respectively. At this point, the capacitance of the EC-cell is unknown, but assuming that it is indeed 9.5 µF, the consequence of neglecting the Steps 1 to 5 of this procedure is an underestimation of 39%, which is remarkably high. In addition to this, Figure 7b shows a data set collected with the EC-cell using all measurement settings optimized except for the measurement window. In the case of the fully optimized data set, a measurement window of 500 mV was used, while in this other case, a much narrower window of 100 mV was used. The average $C_{DL}$ obtained *via* linear fit was 6.0 µF, corresponding to a deviation of 36.8%. This observation demonstrates that the first step of the procedure, that is, the careful selection of a suitable measuring window, has a much larger impact on the outcome of the analysis than the optimization of other measuring settings shown herein.

**Summary of Step 6**: Select a model to fit the experimental data and extract the double layer capacitance. It is recommended to use the allometric fit as a starting model (see Step 7).

*Step 7: Evaluate validity of analysis*

The last step of the procedure, Step 7, focuses on the evaluation of the validity of the result obtained in Step 6. Note that Steps 6 and 7 are meant to be iterative (see Figure 1d) until the validity of the outcome is demonstrated. One approach to achieve this consists in cross-checking the result with a different technique, if possible. In the case of the determination of $C_{DL}$ a technique applicable for the two systems investigated here is EIS. Nyquist plots of the RC-circuit and of the EC-cell are shown in **Figure 1c** and **Figure 8**, respectively. The plots were fitted using models corresponding to an ideal RC circuit (**Figure 1a**) and to a constant phase element (CPE) circuit (**Figure 8**, inset), respectively. In the case of the RC-circuit, a capacitance of 19.8 µF was obtained, which is in full agreement with the result obtained from the voltammetric method, strongly suggesting that the data analysis (Step 6) was adequately conducted. However, this is not the case for the EC-cell. The capacitance obtained from EIS was found to be 14.1 µF, as opposed to 9.5 µF observed with the (most carefully conducted) voltammetric method, corresponding to a deviation of 32.6% with respect to the $C_{DL}$ value obtained by EIS. This discrepancy confirms that the ideal capacitor model is not suitable for describing the behavior of the EC-cell, as discussed in previous sections (Step 5 and 6). Moreover, the need of considering a CPE to model the EIS data, which is typically found for non-ideal capacitive behavior, supports further this observation [39, 40].

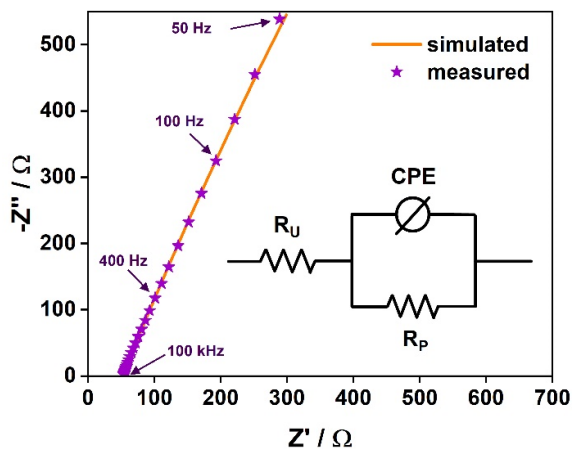

**Figure 8.** Nyquist plot of the EC-cell simulated and measured by electrochemical impedance spectroscopy in the frequency range from 100 kHz to 50 Hz at open circuit potential with an amplitude of 10 mV (RMS). The inset shows the equivalent circuit used for the simulation ($R_U$=52.22 Ω, $R_P$=13.18 kΩ, Yo=21.21x10$^{-6}$, a=0.7564).

It should be mention that using EIS for the determination of $C_{DL}$ comes with difficulties analogue to those of the voltammetric method. On the one hand, electrochemical systems often require the use of EIS models based on more complex equivalent systems [30, 41–43], which may challenge their interpretation and thus the extraction of the $C_{DL}$ [40, 44]. Moreover, the validity of the obtained $C_{DL}$ value also depends on understanding instrumental limitations, as well as on carefully selecting the measurement settings, model and its parameters [45]. In the case of the EC-cell, the CPE model describes fairly the measured data points within the frequency range from 100 kHz to 50 Hz, whereas lower frequency points deviate substantially from the selected model, indicating the presence of additional circuit elements (**Figure S3**). On the other hand, EIS is a technique considerably less accessible than cyclic voltammetry. In the cases where the outcome of the voltammetric method cannot be verified by an alternative method, validation of the analysis can be achieved by an evaluation of the goodness-of-fit parameters with the aim of assessing whether the observed data agree with the predictions of the used regression model. The coefficient of determination, also known as R squared ($R^2$, $0 > R^2 > 1$), is a goodness-of-fit parameter of common use, where a larger $R^2$ is understood as a better fit as it measures the extent to which the variation of the dependent variable (charging current) is explained by the independent variable (scan rate) [46]. The difficulty of its interpretation lies in defining a threshold, i.e., the minimum $R^2$ value for which the validity of a fitting model will be unarguably demonstrated. **Table 1** summarizes the $C_{DL}$ and $R^2$ values corresponding to the linear regression fits shown in **Figure 7** for the two investigated systems. Interestingly, all $R^2$ values observed were relatively high, with the lowest $R^2$ value of 0.9735 corresponding to data obtained for the EC-cell system using non-optimized measuring settings, and the highest $R^2$ value of 0.99999 in the case of the RC-circuit, with the experimental data obtained in a fully optimized manner. Moreover, the data set corresponding to the RC-circuit acquired without optimization of the measurement settings delivered $R^2$ values above 0.999, indicating that, while the fitting model seemed to be suitable for the analysis of the data, such high $R^2$ value cannot be interpreted as a validation of the experimental procedure. In other words, selecting a suitable model leads to high $R^2$ independently of whether the data was acquired correctly or not. Consequently, largely deviating $C_{DL}$ values can be obtained along with high $R^2$ values. Moreover, it is important to consider that $R^2$ does not discriminate reliable from abnormal data, nor delivers information about the distribution of the data points, as illustrated clearly by the Anscombe's quartet [47]. This observation, on the one hand, enhances the importance of observing graphically the collected data as well as selecting carefully the datapoints suitable for analysis (Step 5), and on the other hand, renders $R^2$ insufficient by itself for evaluation of the overall analysis.

A parameter that can be used for assessing the suitability of a linear model is known as the percentage of non-linearity (%nL) [48], and its maximum value within a data set can be obtained according to **Equation 6**. In the case of the determination of $C_{DL}$, $\Delta Y_{max}$ is the maximum difference between the current measured and the current calculated from the equation resulting from the linear regression, while $Y_{max}-Y_{min}$ represents the difference between the maximum and minimum current value measured.

$$\%nL_{max} = \frac{|\Delta Y_{max}|}{|Y_{max}-Y_{min}|} \cdot 100 \qquad (Eq.\ 6)$$

It has been proposed that obtaining %nL lower than 2.5% can be considered a fair indicator that the linear model used is suitable for the fitted data set [48]. Considering this, $\%nL_{max}$ were determined for the different data sets evaluated herein (Table 1) and it was found that the condition $\%nL_{max} > 2.5\%$ was only met by the sets corresponding to the RC-circuit, regardless of whether the measurement parameters were optimized or not. The convenience of the non-linearity parameter is that it allows the observation of deviations from the linear model which can be easily overlooked based on $R^2$ values as well as on the appearance of the current vs scan rate plots when displayed with large X-axis scales. For instance, **Figure 7b** is displayed with a current scale of twice the maximum of the anodic current measured, giving the false appearance of a small deviation of the data points with respect to the linear fit. Thereby, it is strongly recommended to make use of this parameter to evaluate the goodness-of-fit where linear regressions are used as fitting model.

From these results (further supported by EIS) it is clear that the linear model of an ideal capacitor is not suitable for describing the EC-cell. Coming back to Step 6, an alternative model was selected (**Equation 7**). The model corresponds to an allometric regression with the variables Y and X as the current and the scan rate, respectively, the slope b as the capacitance and an exponent that compensates deviations from linearity.

$$Y = bX^\alpha \qquad (Eq.\ 7)$$

This model is analogous to a model used for redox systems in which the relation of $i_{peak}$ vs $v^\alpha$ (peak current vs scan rate to the power of the exponent α) is investigated in relation to the electron transport process. The exponent takes values from 0.5 to 1, generally associated with diffusion-control and surfaced confined electron transport, respectively [49]. In the case of the determination of $C_{DL}$, when the exponent α is equal to 1, the model becomes that of the ideal capacitor (Equation 5), and thus α may be interpreted as a measure of the deviation from the ideal linear behavior. **Figures 9a** and **9b** show the current as a function of the scan rate extracted from the anodic scans of CVs collected with fully optimized measurement settings for the RC-circuit and the EC-cell, respectively. Additionally, a comparison between linear and allometric fits are shown in the same Figure. The resulting $C_{DL}$ and α values obtained for both the anodic and cathodic scans are summarized in **Table 2**.

In the case of the RC-circuit, the resulting allometric and linear fits overlap with each other, as opposed to the EC-cell where not only the two resulting fits do not overlap, but also it is clear that the allometric model better fit the measured data. The average $C_{DL}$ was found to be 19.99 μF with exponents α close to 1 (0.997 and 0.995 for the anodic and cathodic data sets, respectively) suggesting indeed an ideal-like capacitive behavior in agreement with previous observations. Moreover, the observed capacitance value represents a deviation of only 0.05% with respect to the nominal value of 20 μF.

The average $C_{DL}$ value obtained for the EC-cell by means of an allometric model was 14.97 μF with α = 0.776 for both the anodic and cathodic data sets, indicating a large deviation with respect to the ideal capacitor model. Interestingly, the value of the exponent α was close to that of the exponent obtained for the CPE element in the EIS simulation shown in Figure 8, which had a value a = 0.7564. The obtained $C_{DL}$ was considerably larger than the value obtained by means of a linear fit model (9.49 μF), and even larger than $C_{DL}$ obtained without having optimized the measurement settings (5.79 μF). Assuming that the $C_{DL}$ value of 14.97 μF was determined correctly, the consequence of having used the ideal capacitor model is an underestimation of capacitance of 36.6% in the best case, and as large as 61.3% when measurement settings optimization was not carefully done.

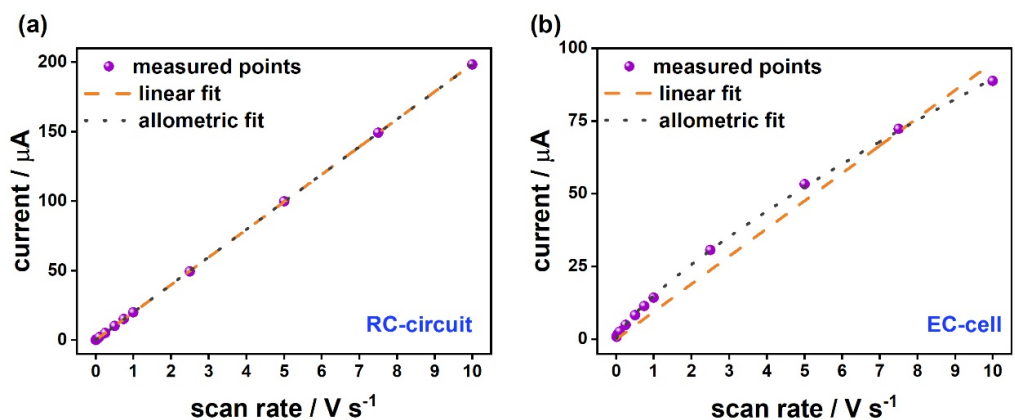

**Figure 9.** Current as a function of scan rate of (a) RC-circuit and (b) EC-cell, obtained from anodic voltammetric scans at different scan rates showing their corresponding linear and allometric fits.

**Table 1.** Double layer capacitance and linear fit parameters of RC-circuit and EC-cell comparing sets of data collected with optimized, non-optimized and partially optimized measurement settings.

| System | Set | $C_{DL}$ / μF anodic | $R^2$ anodic | %$nL_{max}$ anodic | $C_{DL}$ / μF cathodic | $R^2$ cathodic | %$nL_{max}$ cathodic | $C_{DL}$ / μF average |
|---|---|---|---|---|---|---|---|---|
| RC-circuit | 1 | 19.87 | 0.99999 | 0.25 | 19.80 | 0.99999 | 0.41 | 19.84 |
| RC-circuit | 2 | 18.90 | 0.9997 | 2.37 | 18.61 | 0.99949 | 1.93 | 18.76 |
| EC-cell | 1 | 9.51 | 0.98979 | 9.10 | 9.46 | 0.9897 | 9.11 | 9.49 |
| EC-cell | 3 | 6.04 | 0.98515 | 9.62 | 5.97 | 0.98555 | 9.34 | 6.01 |
| EC-cell | 2 | 5.79 | 0.9735 | 12.85 | 5.56 | 0.98026 | 11.77 | 5.68 |

1. Fully optimized data set, including scan rate range and measurement window
2. Non-optimized data set
3. Optimized data set, excluding measurement window (100 mV window used)

To evaluate the validity of the new results (back to Step 7), one can resort to a comparison of $R^2$ obtained with the linear (reported in Table 1) and allometric models (reported in Table 2) along with the comparisons shown in Figure 9. In the case of the RC-circuit, using an allometric or a linear fit did not show a difference in terms of $R^2$, which was above 0.9999 in both cases. Contrariwise, $R^2$ observed for the EC-cell with the allometric model were higher (above 0.999) than those observed with the linear model (below 0.990), indicating once more that the former was indeed a more suitable model than the latter for this particular electrochemical system. $C_{DL}$ values obtained by this model resulted in a deviation of 6.4% with respect to the value obtained by EIS (**Figure 9**, Table 2) as opposed to the deviation of 32.6% observed with the linear fit model, further supporting the suitability of the allometric model.

**Summary of Step 7**: Evaluate carefully the validity of the data analysis by cross-checking with a different method or by means of goodness-of-fit parameters. Be aware of the limitations of either options and critical with their interpretation. If necessary, look for a more suitable model.

Finally, $C_{DL}$ values obtained with the RC-circuit and with the EC-cell using optimized and non-optimized measurement parameters fitted to either the linear or the allometric model are shown in **Figure 10**, illustrating clearly that neglecting the optimization of measurement settings, the adequate selection of a suitable model for the investigated system, or both, leads to substantial deviations of the double layer capacitance. Thus, we stress the importance of following carefully and critically each of the considerations discussed in the seven-step procedure proposed herein.

Table 2. Double layer capacitance ($C_{DL}$) and corresponding allometric fit parameters of RC-circuit and EC-cell with data sets collected with optimized measurement settings. $C_{DL}$ obtained by EIS is shown for comparison.

| System | $C_{DL}$ / μF anodic | α anodic | $R^2$ anodic | $C_{DL}$ / μF cathodic | α cathodic | $R^2$ cathodic | $C_{DL}$ / μF average | EIS α | EIS $C_{DL}$ / μF |
|---|---|---|---|---|---|---|---|---|---|
| RC-circuit | 19.98 | 0.9974 | 0.99999 | 20.01 | 0.9946 | 0.9999 | 19.99 | - | 19.75 |
| EC-cell | 15.01 | 0.7760 | 0.9995 | 14.93 | 0.7760 | 0.99934 | 14.97 | 0.7564 | 14.07 |

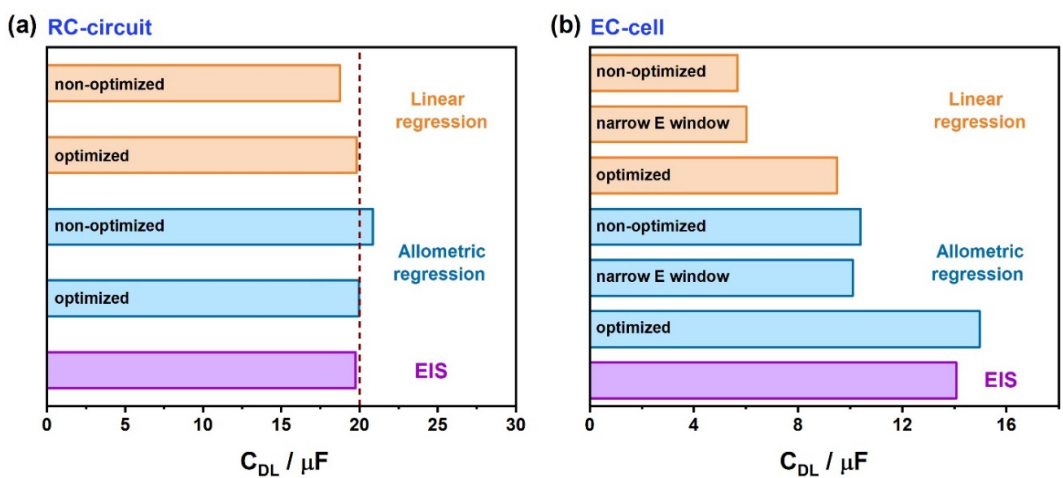

**Figure 10.** Double layer capacitance values obtained with the (a) RC-circuit and (b) the EC-cell. The dash line indicates the nominal capacitance of the RC-circuit. Values obtained from scan rate-dependent cyclic voltammetry using linear and allometric regressions are shown in orange and blue, respectively. Values obtained from electrochemical impedance spectroscopy are shown in purple. Text in the bars indicate whether the data acquisition was done using optimized or non-optimized measuring settings. In the case of the EC-cell the case where all measuring parameters were optimized except for the potential window is indicated as "narrow E window".

### 4. Final Remark: Each μF counts!

Throughout this work, it has been shown that different aspects related to data acquisition as well as data processing must be taken into consideration when conducting quantitative voltammetric analyses, and it has been demonstrated that, when neglected, they can easily lead to substantial deviations. In the case of the determination of $C_{DL}$, these deviations frequently show themselves as an underestimation of its value, which thereby leads to an underestimation of the ECSA (Equation 1). If ECSA is then used to determine current density (j) according to **Equation 8**, the resulting value of j will be overestimated.

$$j = \frac{i}{ECSA} \quad \text{(Eq. 8)}$$

The consequence of overestimating the current density is that it overestimates catalytic activity. The question is then by how much? To try to answer this question, let us assume an underestimation of $\Delta C_{DL} = 1$ μF.

Considering the $C_S$ value of 40 µF cm$^{-2}$, which is often used (though rarely justified) in alkaline media [4], the resulting ECSA would be underestimated by 0.025 cm$^2$, which is equivalent to–apparently–having an additional flat electrode of 1.8 mm diameter per µF underestimated. It is worth noting that the validity of using material-independent $C_S$ values for the calculation of ECSA is a topic that deserves deep revision and discussion, nonetheless as it is out of the scope of the present work, it will not be discussed further herein. Yet, we may use lower or higher $C_S$ values as way of example: with a $C_S$ of 20 and 80 µF cm$^{-2}$, the resulting ECSA is underestimated by 0.05 and 0.0125 cm$^2$, respectively, equivalent to having additional flat electrodes of 2.5 and 1.3 mm, respectively, per µF underestimated.

To illustrate what is the impact of underestimation of ECSA due to $\Delta C_{DL}$, let us take the case of the oxygen evolution reaction. A common metric to describe and compare the catalytic activity of oxygen-evolving materials is the potential ($E_{OER}$) at which certain current density ($j_{OER}$), typically 10 mA cm$^{-2}$, is attained (**Figure 11a**). Underestimations of ECSA of 0.0125, 0.025 and 0.05 cm$^2$ lead to mistakenly reporting $E_{OER}$ at $j_{OER}$ values of 9.88, 9.76 and 9.52 mA cm$^{-2}$, respectively, instead of at 10 mA cm$^{-2}$, thus mistakenly delivering an apparent higher activity (lower overpotential), and this occurs as a consequence of an underestimation of only 1 µF. The variations of $j_{OER}$ with underestimations of $C_{DL}$ of 1, 2, 5 and 10 µF are shown in **Figure 11b** calculated with $C_S$ values between 10 and 130 µF cm$^{-2}$, which are values reported typically in the literature [4]. The larger $\Delta C_{DL}$, the larger the deviation of $j_{OER}$ with respect to the 10 mA cm$^{-2}$ value at all $C_S$ values, impacting much more substantially materials with lower $C_S$ values.

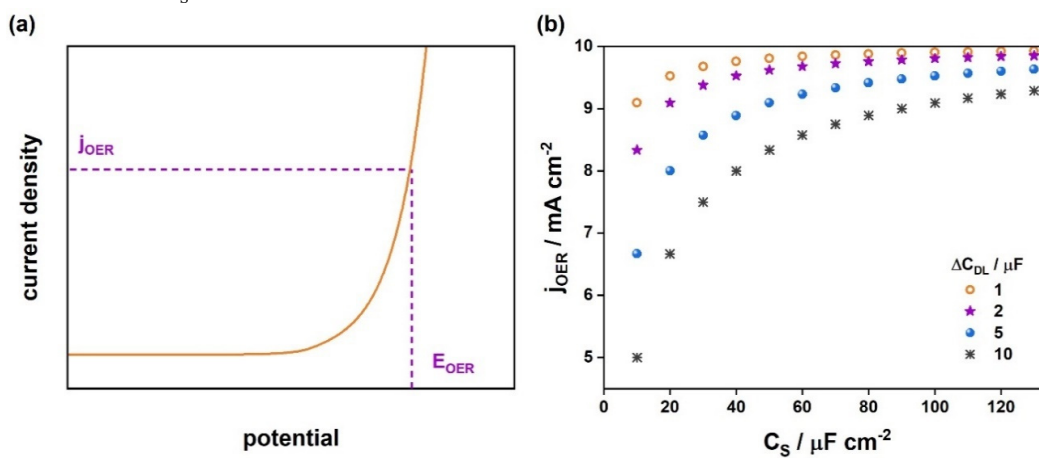

**Figure 11**. (a) Representation of activity metrics $j_{OER}$ and $E_{OER}$ extracted from a simulated polarization curve corresponding to the oxygen evolution reaction. (b) Variation of the metric $j_{OER}$ as a function of the specific capacitance ($C_S$) observed with double layer capacitance underestimations ($\Delta C_{DL}$) of 1, 2, 5 and 10 µF. $j_{OER}$ was calculated assuming that $\Delta C_{DL} = 0$ results in $j_{OER} = 10$ mA cm$^{-2}$.

Clearly, $C_{DL}$ is crucial for an adequate assessment of the catalytic activity when it is done by means of current density-based activity metrics, and the accuracy with which it is determined must be carefully observed. The seven-step procedure proposed herein proves to be of great relevance in the determination of $C_{DL}$ since it effectively leads to a higher probability of performing accurate voltammetric measurements as well as valid analyses. Because of that, the procedure is not limited to the determination of $C_{DL}$, but it can be readily applied to any other voltammetry-based analysis aiming to deliver quantitative results.

## 5. Conclusion

In the context of characterization of electrochemical properties of catalytic materials, we presented a procedure which seeks to increase the overall quality of voltammetric experiments, in particular for the determination of double layer capacitance ($C_{DL}$) by scan rate-dependent cyclic voltammetry. The procedure is divided in seven steps, each of which focuses in a specific aspect related to data acquisition, data analysis or validation of the analysis. The first three steps are dedicated to the selection and optimization of various measurement settings, including the potential window, dynamic compensation of $iR_U$-drop, current range, and data sampling mode within the voltage steps of a voltammetric staircase. To observe the impact of these parameters on the determination of $C_{DL}$, two systems were studied: a resistor-capacitor electric circuit and a glassy carbon disk in an electrochemical cell. It was demonstrated that the first step of the procedure, that is, the careful selection of a suitable potential window, has a much larger impact on the outcome of the analysis than the optimization of the other measuring settings investigated herein. In the case of the studied electrochemical system, the use of a 100 mV potential window, which is a range of potential used typically for these analyses, proved to be too narrow leading to an underestimation of $C_{DL}$ of about 36% with respect to the case where the potential window was optimized (500 mV). The fourth and fifth steps of the procedure focus on the collection and selection of experimental data, respectively, considering a careful observation of the variation of charging current with respect to the scan rate within a wide scan rate range. The sixth and seventh steps of the procedure focus on the selection of a fitting model and the evaluation of its suitability for the determination of $C_{DL}$, respectively. Two models were compared to fit the data acquired for both the electric circuit and the electrochemical system: a linear model based on the ideal capacitor behavior, which is of widespread use within the research community for this purpose, and an allometric fit, which we proposed herein as an alternative model. By evaluating the selected models using goodness-of-fit parameters, including $R^2$, maximum non-linearity, and normalized $R^2$, it was demonstrated that while the linear model was suitable to describe the behavior of the electric circuit, it delivered largely deviated $C_{DL}$ values for the electrochemical system, whereas the allometric model was suitable to fit the data of both systems, proving useful for both ideal and non-ideal systems. In addition to this, we showed that high $R^2$ values were observed for all data sets, including those collected with non-optimized measuring parameters fitted to the linear model, indicating that $R^2$ by itself does not suffice to validate the outcome of these analyses. We further supported these observations by comparison with $C_{DL}$ obtained by electrochemical impedance spectroscopy.

In summary, it was shown that the probability of determining $C_{DL}$ in a successful manner is effectively increased by carefully following the seven-step procedure proposed herein, since it is built up as a series of strategic steps aiming to circumvent sources of error. Moreover, the seven-step procedure is not limited to the determination of $C_{DL}$, but can be (and should be) applied to any other voltammetry-based analysis that aims to deliver quantitative results. We stress the importance of regarding the proposed methodology as a guideline that should be adapted to the characteristics of the particular electrochemical system studied and not conversely. It is important to consider carefully all the components of the electrochemical system as well as the potentiostat capabilities in order to select the suitable techniques, conditions, and measuring settings for its study. Finally, it is strongly recommended to make use of dummy cells (or electric circuits) to fully understand the potentiostat capabilities and its limitations, and use it accordingly for subsequent measurements.

**Acknowledgements**

Authors are grateful to Mr. Hans-Jürgen Wittmaack from the Elektronik-Werkstatt at Helmholtz-Zentrum Berlin for building the electric circuits used. We thank Krystian Lankauf (Gdańsk University of Technology) for bringing the percentage of non-linearity to our attention. This project has received funding from the European Research Council (ERC) under the European Union's Horizon 2020 research and innovation programme under grant agreement No 804092.

# Supporting Information

# Seven steps to reliable cyclic voltammetry measurements for the determination of double layer capacitance

**Dulce M. Morales[1] and Marcel Risch[1]**

[1] Nachwuchsgruppe Gestaltung des Sauerstoffentwicklungsmechanismus, Helmholtz-Zentrum Berlin für Materialien und Energie GmbH, Hahn-Meitner-Platz 1, 14109 Berlin, Germany

E-mail: marcel.risch@helmholtz-berlin.de

1. Sampling mode for data acquisition

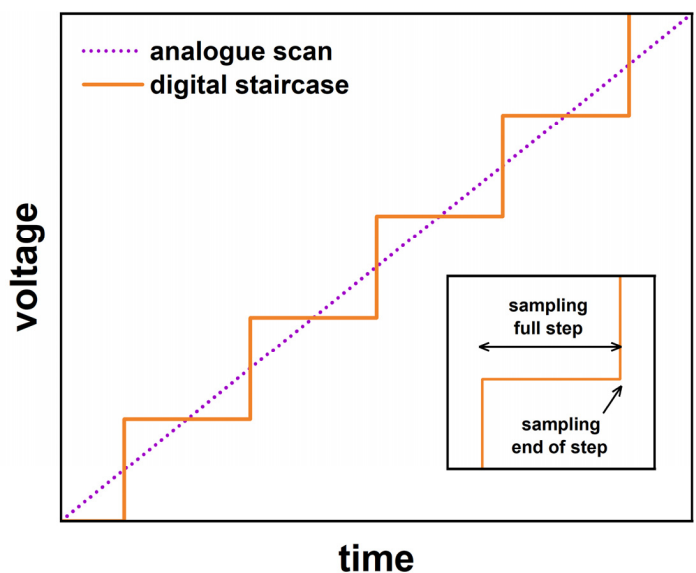

**Figure S1.** Comparison between an analogue potential scan and a digital staircase. Inset illustrates data acquisition conducted by sampling the full potential step and by sampling at the end of the step. Figure adapted from [1] with permission from Gamry Instruments Inc.

## 2. Randles circuit model

An electrict circuit was built to emulate a Randles cell. It consisted of a 100 µF aluminum electrolytic capacitor (± 20% tolerance, rated voltage 35 V, Frolyt) parallel to a multiturn cermet trimming potentiometer with maximum resistance of 500 Ω (± 10% tolerance, TT Electronics), both soldered in series with a 50 Ω metal film resistor (± 1% tolerance, Yageo). A photograph and a diagram of the circuit are shown in **Figure S2a** and **S2b**, respectively. The resistances, built in series and in parallel with respect the capacitor, are here denoted as $R_S$ and $R_P$, respectively.

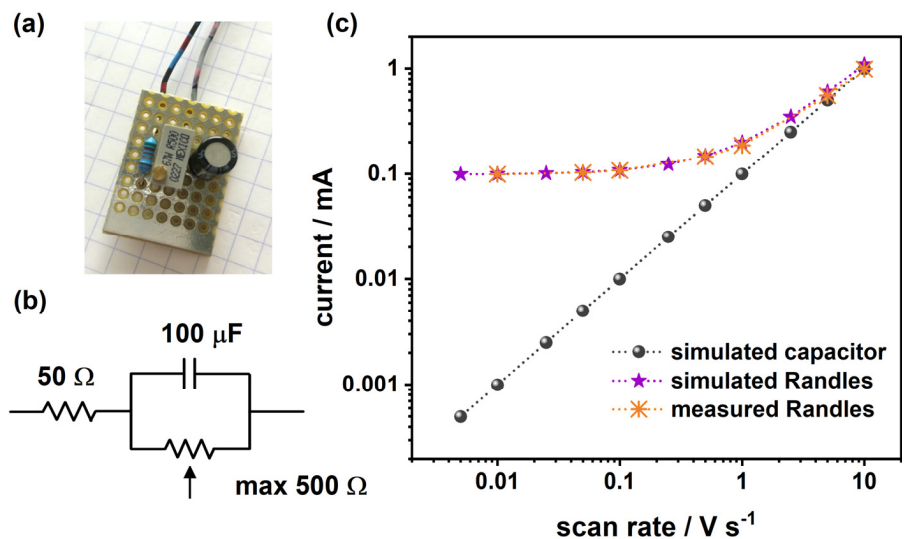

**Figure S2**. (a) Photograph and (b) electrical diagram corresponding to the Randles circuit used in this work. (c) Current as function of scan rate extracted from cyclic voltammograms recorded with the built Randles circuit with a dynamic compensation of 45 Ω, compared with simulated data corresponding to an ideal Randles circuit (500 Ω resistor parallel to 100 µF capacitor, both in series with a 50 Ω resistor) and an ideal resistor-capacitor circuit (50 Ω resistor in series with 100 µF capacitor).

Using the seven-step procedure described in the main manuscript, cyclic voltammograms were recorded in the range from 0 to 0.1 V at different scan rates with $R_P$ set to 500 Ω. The data corresponding to current measured during anodic scans was extracted from the recorded voltammograms at a voltage of 0.05 V and it is shown as a function of the scan rate in **Figure S2c** (orange). The resulting curve is compared to data current vs scan rate (i vs v) simulated from two different models: an ideal capacitor (**Equation 1**), and an ideal Randles circuit (**Equation 2**) considering the voltage applied (V) and the total resistance ($R_T$) of the circuit. $R_T$ was calculated according to **Equation 3**, considering $R_P$ and $R_S$, as well the resistance compensated dynamically ($R_C$). The comparison of the three data sets shows on the one hand, that the model indicated by Equation 2 describes appropriately the data measured, and on the other hand, that only a high scan rates, where the capacitive contribution to the net current is considerably large, the behaviour of the Randles circuit resembles that of the ideal capacitor.

$$i = v \cdot C = v \cdot (100 \text{ µF}) \qquad \text{(Eq. 1)}$$

$$i = \frac{V}{R_T} + v \cdot C = \frac{(0.05 \text{ V})}{(500 \text{ Ω})} + v \cdot (100 \text{ µF}) \qquad \text{(Eq. 2)}$$

$$R_T = R_P + R_S - R_C = (500 \text{ Ω}) + (50 \text{ Ω}) - (45 \text{ Ω}) = 105 \text{ Ω} \qquad \text{(Eq. 3)}$$

3. Electrochemical impedance spectroscopy

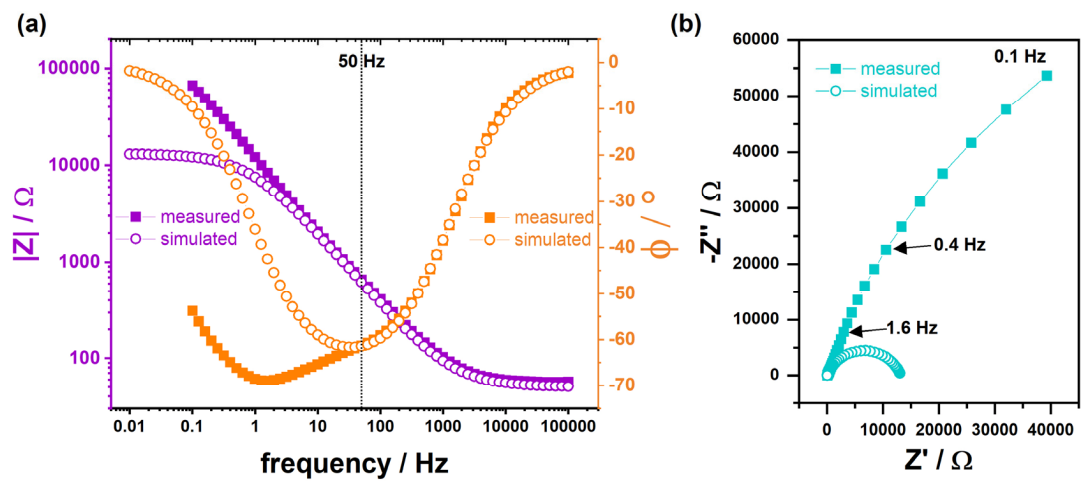

**Figure S3**. (a) Bode plot and (b) Nyquist plot of the EC-cell (O) simulated and (■) measured by electrochemical impedance spectroscopy in the frequency range from 100 kHz to 0.1 Hz at open circuit potential with an amplitude of 10 mV (RMS). Simulation parameters are: $R_U$=52.22 Ω, $R_P$=13.18 kΩ, $Y_o$=21.21x10$^{-6}$, a=0.7564.

**References**

[1] Gamry Instruments *Measuring surface related currents using digital staircase voltammetry (Application note)* Retrieved from: www.gamry.com/application-notes/physechem/cyclic-voltammetry-measuring-surface-related-currents/ (accessed 17 Sep 2020)